\documentclass[electronic]{vgtc}             

\ifpdf
  \pdfoutput=1\relax                   
  \usepackage{graphicx}                
  \DeclareGraphicsExtensions{.pdf,.png,.jpg,.jpeg} 
\else
  \usepackage{graphicx}                
  \DeclareGraphicsExtensions{.eps}     
\fi%

\graphicspath{{figures/}{pictures/}{images/}{./}} 

\usepackage{microtype}                 
\PassOptionsToPackage{warn}{textcomp}  
\usepackage{textcomp}                  
\usepackage{mathptmx}                  
\usepackage{amsfonts}
\usepackage{times}                     
\usepackage{cite}                      
\usepackage{tabu}                      
\usepackage{booktabs}                  
\usepackage{csvsimple}

\usepackage{mathtools, nccmath}

\usepackage[usenames,dvipsnames,svgnames,table]{xcolor} 
\definecolor{ngray}{rgb}{0.5,0.5,0.5}
\definecolor{lgray}{rgb}{0.85,0.85,0.85}
\usepackage{listings,multicol}
\lstdefinelanguage{Diderot}{%
  morekeywords={%
    bool,%
    die,%
    else,%
    false,field,foreach,sphere,%
    const, inv,
    identity,if,image,load,in,initially,input,int,%
    kernel,%
    nan,new,%
    output,%
    real,%
    return,
    overload,
    refCell,
    position,
    function,stabilize,strand,string,%
    inside,normalize,tensor,true,%
    update,type,%
    vec2,vec3,vec4,%
    zeros},
  sensitive,%
  morecomment=[s]{/*}{*/},%
  morecomment=[l]//,
  morestring=[b]"}%

\onlineid{0}

\vgtccategory{Research}

\vgtcinsertpkg

\title{
Point Movement in a DSL for Higher-Order FEM Visualization
}
\author{Teodoro Collin\textsuperscript{1}\\ %
 \and Charisee Chiw\textsuperscript{2}\\ %
 \and L. Ridgway Scott\textsuperscript{1}\\
 \and John Reppy\textsuperscript{1}\\
 \and Gordon Kindlmann\textsuperscript{1}\\
}

\affiliation{ \scriptsize \textsuperscript{1}Department of Computer Science, University of Chicago\thanks{e-mails:\{teocollin,ridg,jhr,glk\}@cs.uchicago.edu} \hspace{2cm} \scriptsize \textsuperscript{2}Galois, Inc.\thanks{e-mail: chiw@galois.com}}

\abstract{
Scientific visualization tools tend to be flexible in some ways (e.g.,
for exploring isovalues) while restricted in other ways,
such as working only on regular grids, or only on unstructured meshes
(as used in the finite element method, FEM). 
Our work seeks to expose the common structure of visualization
methods, apart from the specifics of how the fields being visualized are
formed.
Recognizing that previous approaches to FEM visualization depend on
efficiently updating computed positions within a mesh, we took an
existing visualization domain-specific language, and added a mesh position type and
associated arithmetic operators. 
These are orthogonal to the visualization method itself, so
existing programs for visualizing regular grid data work,
with minimal changes, on higher-order FEM data.
We reproduce the efficiency gains of an earlier \emph{guided search}
method of mesh position update for computing streamlines, and we
demonstrate a novel ability to uniformly sample ridge surfaces of
higher-order FEM solutions defined on curved meshes.
}

\CCScatlist{ 
  \CCScatTwelve{Software and its engineering}{Software notations and tools}{Context specific languages}{DSLs}
}

\begin{document}
\firstsection{Introduction} 

\maketitle

\label{sec:intro}


Novel methods of high-performance and scalable scientific
visualization typically support interactively exploring various parameters
(e.g., volume rendering transfer functions, or streamline
seedpoints), while constraining the \emph{form} of data being visualized.
That is, tools for visualizing large biomedical imaging volumes are sensibly
specialized for the regular grids that such data is acquired
on, just as fluid flow visualization tools are
specialized for the finite element method (FEM) meshes on which those
phenomena are simulated.
Yet from the high-level mathematical standpoint of either characterizing
existing visualization methods, or exploring the value
of new ones, the specialization of tools
to data forms is unfortunate: volume rendering is a kind of
integration, and streamlines are solutions to ODEs, regardless of
how exactly scalar or vector fields are defined on a grid or mesh.
%
%
%
Visualization research may benefit from systems that can take a
high-level specifications of a visualization method, and a separate
description of how data and fields are formed, and then compile
programs that both run efficiently on the given data and
support exploration of the relevant parameter spaces.

Domain-specific languages (DSLs) for scientific visualization
partially address this need by specializing for a class of algorithms
and one form of data: regular grids~\cite{GLK:McCormick07,shadie-dsl,GLK:pldi12,GLK:McCormick2014,GLK:vislang,GLK:vivaldi}.
%
In their own ways, these languages work to separate the legible
expression of visualization algorithms from technical details of
data access or parallel computing.
Our current work, however, explores whether a scientific visualization
DSL can also be general with respect to data form, so that a
program that works efficiently for data on a regular grid
can, with minimal changes, also work on an unstructured mesh.
Particularly challenging are higher-order finite element simulations, which use
higher-order polynomials in both the geometry of mesh
elements (so they can be curved) and the form of solutions within
elements (beyond affine functions), since this increases
the implementation complexity of mere field evaluation.

We present \emph{preliminary} work on extending the open-source compiler for a
scientific visualization
DSL~\cite{GLK:pldi12,GLK:vis15,GLK:eurovis18}, previously limited to
regular grids, to also work on higher-order FEM data.
Our long-term goal is to connect previous
FEM visualization methods~\cite{coppola_nonlinear_2001,
nelson_ray-tracing_2006,meyer_particle_2007,
nelson_gpu-based_2011,pagot_efficient_2011,pagot_interactive_nodate,
nelson_elvis:_2012,nelson_gpu-based_2014,
jallepalli_treatment_2018} by simplifying how
they can be expressed and combined in working code.
Our current focus is just two kinds of visualizations, both involving
computations on a discrete set of points: streamlines~\cite{coppola_nonlinear_2001}, and particle
systems sampling surface features~\cite{meyer_particle_2007}.
The inner loops of both methods share an essential element,
\emph{point movement}, i.e., incrementing a position within a higher-order FEM
domain by some update vector.
We seek a programming language that allows clearly legible implementations
of point movement (as a specification of an aspect of the data form),
and of the visualization method itself, to be combined in a single program.

Our main contribution approaches this by demonstrating how
adding a type for representing FEM mesh positions to a DSL,
and overloading operators on that type, simplifies
implementing visualization methods that rely on point movement.
A supporting contribution, ridge surface extraction in a curved
finite element mesh by a particle system, uses the main contribution,
and also suggests how other new visualizations could be created by
combining general methods with specializations to data form.
We hope our work (which itself will be made open-source available)
can eventually lower the implementation cost of FEM visualization, as well as help extend
standard visualization algorithms to other more general
domains (e.g. manifolds) in which points and vectors have
distinct significance.

\section{Finite Element Method (FEM) Background}
\label{sec:FEMb}

A brief explanation of a \emph{simplified} and \emph{typical} use of FEM
will support a description of our work.
For a solution to a partial differential equation (PDE) $u:\Omega\to\mathbb{R}^{n}$,
FEM uses a finite dimensional vector space of
functions (function space) $V$ to find an approximate
solution, $u_{V}\in V$, to $u$.
The space $V$ is created by discretizing the world-space domain $\Omega$
into a collection of disjoint cells ${\{K_{i}\}}$,
defining a function space $P_{i}$ on each $K_{i}$, and
then combining all the $P_{i}$, e.g., $v\in V$ if and only if
$v_{\mid K_{i}}\in P_{i}$ for all element indices $i$~\cite{Brenner2008}.
The per-element function space is typically
$P_{i} = \{p\circ T_{i}^{-1} \mid p\in P\}$, where $P$ is a
finite dimensional function space
on a convex polytope $K$ (the reference cell),
each $T_{i}$ is an injective $C^{\infty}$ mapping
from $K$ to world-space,
and both $P$ and $T_{i}$ are polynomial.
Consequently, finite element solvers do not need to explicitly
represent the $P_{i}$ (or compute $T_{i}^{-1}$), and can
compute all quantities on the reference cell $K$~\cite{Brenner2008,Rathgeber2016,cantwell_nektar++:_2015,fenicsbook2015, BangerthHartmannKanschat2007}.

Unfortunately, visualization naturally works in world-space $\Omega$.
Within some $K_{i}\subset\Omega$, the PDE solution being visualized $u_{V}$ will be
represented in a chosen basis $\{p_{j}\}$ for $P$ as
\begin{equation}
  u_{V}(x)_{\mid K_{i}} = \sum_{j} c_{j}p_j(T_{i}^{-1}(x)). \label{eq:uofx}
\end{equation}
Using higher-order FEM, with non-linear $T_{i}$, increases the
computational cost for a \emph{naive} visualization algorithm to traverse
just a single cell $i$, since each of the many evaluations of
$T_{i}^{-1}$ in (\ref{eq:uofx}) require multiple Newton iterations.
Moreover, as the visualization traverses world-space, for each
point $x\in\Omega$ it needs to compute $T_{i}^{-1}$ for
many different cells $i$ in order to find
the $i$ for which $x\in T_{i}(K)$~\cite{nelson_elvis:_2012}.
Visualization algorithms specialized to finite elements
avoid the cost of $T_{i}^{-1}$ by replacing, when possible,
world-space evaluation of the approximate solution $u_{V}$ via (\ref{eq:uofx}) with
\begin{equation}
  f_{i}(x) := \sum_{j} c_{j}p_j(x), \label{eq:f_iofx}
\end{equation}
where $f_{i}\colon K\to\mathbb{R}^{n}$ is the
evaluation of $u_{V}$ on the \emph{reference cell}
with respect to element $i$.
\section{Related Work}
\label{sec:visb}

Many tools for visualizing FEM solutions
(including ParaView~\cite{paraviewBook}, Gmsh~\cite{gmshCite}, and
GLVis~\cite{glvis-tool})
use \emph{tessellation}, i.e.,
approximating one higher-order element with multiple smaller affine
elements~\cite{remacle_ecient_2005}.
This is good for simple visualizations (e.g. colormapping
$u_{V}$), but more problematic for more complicated ones,
such as volume rendering or those that require higher order derivatives~\cite{nelson_elvis:_2012}.
The tessellation framework of Schroeder et al.\ is in principle
general and accurate with respect to visualization
method~\cite{schroeder_methods_2006}, but we are unaware of its
application beyond isosurfaces and streamlines.

Our work follows a different strategy, advanced by Nelson et al.,
which directly visualizes elements, without tessellation.
Via algorithms that directly manipulate $u_{V}, f_{i}, K$ and $T_{i}$,
these authors create fast and accurate methods for ray-tracing isosurfaces,
cut surfaces, and volume rendering~\cite{nelson_ray-tracing_2006,nelson_gpu-based_2011, nelson_gpu-based_2014}.
They also combine methods into ElVis, a GPU-based interactive GUI
application, which offers some generality over data forms via a plugin
architecture that supports a small set of visualization
algorithms~\cite{nelson_elvis:_2012}.
Our DSL, however, allows more room to explore implementation
variation within or between data forms, at the expense of
lower computational performance relative to hand-written low-level code.

A variety of other previous work investigates FEM visualization
under various accuracy, expression, and performance constraints,
as surveyed by Nelson et al.~\cite{nelson_elvis:_2012}.
Some work focused on fast and accurate visualizations of
curved quadratic and cubic elements~\cite{WileyThesis} while other work achieved
interactive volume rendering of higher order elements~\cite{uffinger_interactive_2010}.
Later sections will describe in more detail the work of Coppola et al.~\cite{coppola_nonlinear_2001}
and Meyer et al.~\cite{meyer_particle_2007},
which is most central for our current work.
%
There is no prior work extracting ridge surfaces from FEM data,
but Pagot et al.\ find ridge lines on affine meshes via PVO and new seed finding
and streamline routines~\cite{pagot_efficient_2011}.
Jallepalli et al.'s smoothing of finite element data could usefully
complement the visualization methods we target~\cite{jallepalli_treatment_2018}.


We also build on related work with DSLs.
Being specific to some domain of algorithms, 
DSLs trade reduced flexibility of the language for (in principle) higher human
productivity of writing new programs within that
domain~\cite{GLK:dsl-when-and-how}.
For our purposes we merely note FEM-related DSLs
for formulating and solving PDEs~\cite{UFL,Rathgeber2016,fenicsbook2015,liszt}
as well as DSLs for processing and visualizing
image and volume data~\cite{GLK:McCormick07,shadie-dsl,GLK:pldi12,halide2013,GLK:McCormick2014,GLK:vislang,GLK:vivaldi,halide2016}.
This list does not fairly describe the sophisticated approaches to
high-performance computing~\cite{GLK:McCormick2014} and
computational scheduling~\cite{halide2013,halide2016}.
We build on Diderot, a visualization DSL limited to regular
grids~\cite{GLK:pldi12,GLK:vis15,GLK:eurovis18}, but distinguished by
offering the mathematical abstraction of a $C^{k}$ tensor field.
Our current work extends how Diderot fields are defined to include
FEM, so that existing Diderot programs can be used with
minimal changes, while introducing a new abstraction, a \emph{mesh
  position}, which supports the convenient expression of previous
methods of moving through the geometry of a curved FEM
mesh~\cite{coppola_nonlinear_2001,meyer_particle_2007}.

\section{Methods}
\label{sec:methods}

\subsection{Point Movement via Guided Search}
\label{sec:movement}

Many scientific visualization algorithms enjoy \emph{spatial coherence}:
field evaluation at (world-space) position $x$ will likely be followed
by evaluation at a nearby $x+v$.
As noted in \S\ref{sec:FEMb}, for simple methods,
the computational expense (from finding the cell containing $x$, and
finding $T_i^{-1}(x)$) of naively evaluating $u_{v}(x)$ via
(\ref{eq:uofx}) might be avoided by evaluations in \emph{reference
  space} via (\ref{eq:f_iofx}), and then forward mapping by $T_i$.
Previous work with more complex visualization methods, however,
demonstrates the value of rapidly approximating $x+v$ in a sequence of reference
spaces, a technique we term \emph{point movement}, so that $u_{V}(x+v)$
can be found faster than via naive re-evaluation of (\ref{eq:uofx}).
For streamlines, Coppola et al.\ name their method of point movement
\emph{guided search}~\cite{coppola_nonlinear_2001}, while a similar
method underlies the isosurfacing particles of Meyer et
al.~\cite{meyer_particle_2007}.

\begin{figure}
\vspace*{-1em}
  \includegraphics[width=\columnwidth]{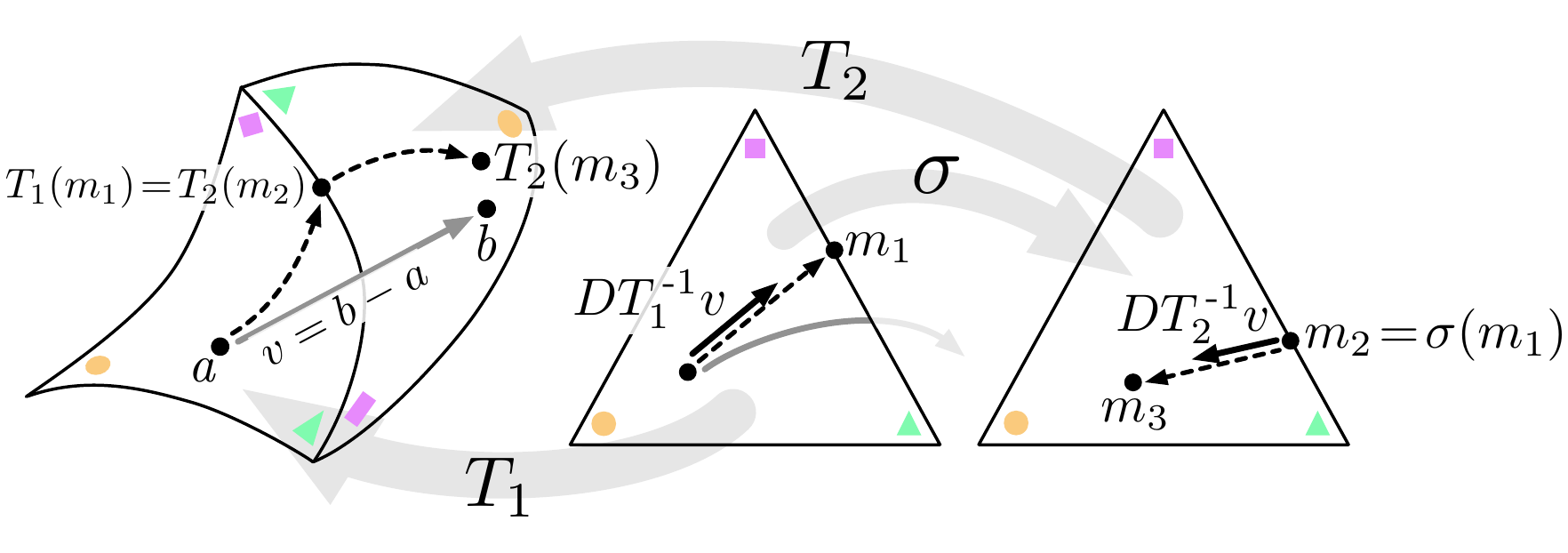}
\vspace*{-2.3em}
\caption{\label{fig:guidepic} Illustration of guided search
  to move $a$ towards $b$, with two
  world-space cells (left) and two copies of reference space (center
  and right), each with transforms $T_1$ and $T_2$. Colored shapes at cell vertices clarify how
  reference spaces connect. At $a$, the velocity $v=b-a$ is transformed by
  $DT_1^{-1}$ to give
  a reference space velocity $DT_1^{-1}v$ along which traversal
  begins.  At cell boundary $m_1$, the permutation $\sigma$ between
  reference vertices determines the
  start $m_2=\sigma(m_1)$ of the next cell traversal, now along
  $DT_2^{-1}v$ at $m_{2}$. More cells may be encountered, until the computed path (shown
  as a dashed line) has run for $\mathrm{time}\!=\!1$, ending at $m_3$.
  The point $T_2(m_3)$ approximates $b$.
}
\end{figure}

Guided search builds on a few technical observations.
First, for world-space position $x\in T_{i}(K)$ and update $v$,
a first-order approximation of the updated position
is $T_{i}^{-1}(x+v) \approx T_{i}^{-1}(x) + (DT^{-1}_{i}(x))v$.
Second, the location where $T_{i}^{-1}(x)+ (DT^{-1}_{i}(x))v$
exits $K$ can be found via geometric computations on
the shape of $K$, common in computer graphics~\cite{RTCD}.
Third, in most meshes, the mapping between reference
positions in two adjacent cells is entirely determined by a
simple permutation $\sigma$ on the vertices of $K$.
Thus for $x\in\partial K$ on the boundary of the reference cell for cells $i$
and $j$, the same world space position is both $T_{j}(\sigma(x))$ and $T_{i}(x)$.
Combining these ideas together yields the guided search algorithm
illustrated in Fig.~\ref{fig:guidepic}.

We make two observations about the context and
implementation of guided search.
Guided search has only been described as a part of a specific
visualization ingredient:
Coppola et al.\ presents guided search as a sub-step
of RK4 integration~\cite{coppola_nonlinear_2001}. 
%
In fact, it can be separated from any particular numerical or visualization method and
framed as a method to update positions by a vector,
a definition in affine geometry~\cite{LinAlgAndGeo}.
%
Second, guided search is complicated enough to warrant
exploring the speed and accuracy of possible variants.
For example, Coppola et al.\ also describe \emph{error-checked guided search},
wherein the search defaults to the naive scheme to
locate $x+v$ if $||T_{i}(T_{i}^{-1}(x) + t(DT^{-1}_{i}(x))v) -
(x+tv)||$ exceeds some threshold.
The same considerations of orthogonality and legibility that
motivate creating DSLs also suggest clearly expressing the point movement
method within the language.

\subsection{FEM Data, Position types, and Overloading}
\label{sec:meshpos}

\begin{figure*}[hbt]
\begin{minipage}{1.0\textwidth}
{\fontsize{6.2}{7.3}\selectfont
\begin{lstlisting}[xleftmargin=1.5em,multicols=2]
input int timeSteps=32;
const int dim = mesh_t.dim;    // (this works in 2 or 3 dimensions)
input real timeEps = 0.0000001;
input mesh_t mesh;
refCell{mesh_t} K = mesh.refcell;
overload position{mesh_t} +(position{mesh_t} x, tensor[dim] delta){
  if (!x.isValid){ return(x);}
  real time = 1;
  position{mesh_t} cmp = x;                   // current mesh pos
  foreach (int i in 0..timeSteps ){           // ("mc"=mesh cell)
    tensor[dim, dim] iJac = inv($\nabla\otimes$(cmp.mc.transform)(cmp.refPos));
    tensor[dim] refDelta = iJac $\bullet$ delta; // reference velocity
    tensor[dim] nPos = cmp.refPos + time*refDelta;
    if (K.isInside(nPos)) {                   // nPos is inside K
      return(cmp.mc.meshPos(nPos));           // x+v=nPos in cmp.mc
    } else { // we left the reference cell; compute when we left
      real eTime = K.exit(cmp, normalize(refDelta));
      time -= time/|refDelta|;         // decrement time remaining
      if (eTime == -1){ // invalid direction, use naive scheme
	return(mesh.findPos(x.worldPos() + delta));
      } // else find the exit location in the next cell
      position{mesh_t} nmp = K.exitPos(cmp, normalize(refDelta));
      if ( !nmp.isValid || time < timeEps){   // left mesh or
	return(nmp);                          // ran out of time
      }
      cmp = nmp;
    }
  } // spent too much time; use naive scheme
  return(mesh.findPos(x.worldPos() + delta));
}

\end{lstlisting}%
}%
\end{minipage}
\vspace{-0.1cm}
\caption{\label{fig:overload}
An overloaded ``\texttt{+}'' operator implements a minimal
version of guided search along with minimal context.
Lines 1-5 declare the necessary inputs.  
Lines 7-9 set up the search. Lines 10-29 are the main body. At each iteration, lines 11 abd 12
transform the change in position via the position's cell's
transform. Lines 14 and 15 return a new position if the transformed
velocity does not take the position outside the cell. Otherwise, lines
17-28 find the time of intersection (line 17), check that the
intersection makes sense (line 19), find the position of this
intersection on the next cell (line 22), check if this is the last
step (line 23), and continue on (line 26). If the loop terminates, line 29 defaults to the naive scheme.} 
\end{figure*}


To demonstrate point movement within a mesh as a programmable and
orthogonal aspect of a FEM visualization algorithm, we augment an
existing scientific visualization DSL with a new position type,
overloaded operators on positions, and the ability to input FEM data.
%
%
We chose the Diderot language because it already simplifies
implementing streamlines and particle systems on regular grids~\cite{GLK:eurovis18,GLK:vis15}, and
because its consistent use of a \emph{field} abstraction facilitates introducing
FEM solutions as a new underlying data form.

Space here permits a high level summary of the language changes required
to create a path from FEM data to existing language objects: domains, fields, and tensors.
The domain of an FEM solution involves a mesh, reference cell domain $K$, and the $T_{i}(K)$;
each requires a new language type 
and constructors via inputs or methods.
Meshes are global (immutable) inputs to the program, supplying
a sequence of cells on the mesh.
The global solution $u_{V}$ can be accessed as a field
after providing a space type, a solution type, and an input.
For fields attached to cells, such as $T_{i}, T_{i}^{-1}$, and $f_{i}$,
the fields are cell methods.
To enjoy the benefits of sampling within a reference cell,
cells provide a \emph{transformed reference field} which
supports evaluating values $f_{i}(x)$ and derivatives $D^{n}(f_{i}\circ T_{i}^{-1})$
at $T_{i}(x)$, so that tensor-valued (world-space) derivatives of $u_{V}$ can be efficiently
sampled from reference space.
Meyer et al.\ also sample gradients and
Hessians from the reference cell, and use a lengthy Einstein notation
derivation to find the world-space derivatives~\cite{meyer_particle_2007}.
All these mechanics are thankfully handled automatically by the Diderot compiler's
internal representation, itself based on Einstein notation~\cite{chiw18}.

To support the notion of a position on a mesh, we added a new position type
that depends on a mesh type; other domain
types could be supported later.
Position values are constructed either with a point in reference space $K$ and
a mesh cell, or via a point in world space; the latter option
corresponds to the naive scheme.
Strands of Diderot computation can be associated with positions,
so that strands (e.g., for particle systems) can query the state
of their neighbors via k-d trees~\cite{diderot:cpc15}.
We also added queries on the reference cell geometry, to
determine when a point leaves its cell by traveling in a direction,
and to learn the corresponding position in the neighboring cell (if it exists).
With all this in place, positions can become
arguments to an overloaded ``\texttt{+}'' operator, the
concise guided search implementation of Figure~\ref{fig:overload}.
%
%
Adding these capabilities to the compiler required adding or changing around
5000 lines of Standard ML, but this cost is once per data form as we can now use the compiler to explore the adoption of many previous Diderot programs to a FEM context that is consistent with \S\ref{sec:FEMb}.
Below, we focus on just two Diderot programs.


The language elements
described above allow separating the expression of visualization
algorithms from both the details of field evaluation
and the details of point movement.
In particular, we were able to modify existing
Diderot programs for streamlines~\cite{GLK:vis15} and particles~\cite{GLK:eurovis18} in regular grids to
work with FEM data via straight-forward transformations.
We declared FEM types and inputs,
added point movement code (Fig.~\ref{fig:overload}),
changed field sampling to a function that samples the reference field using a position,
and changed several types from vectors to positions.
While this seems extensive, besides the copy-pasted parts,
disruption to existing code was minimal:
15 lines changed in a 30-line streamline program,
30 lines changed in an 80-line isosurfacing particle program,
and 40 lines changed in a 300-line program for general
feature sampling with particle systems.
%
%
%
%
%
The full analysis is
in the supplementary materials.


\section{Results}
\label{sec:results}

Our results all use curved meshes with cubic $T_i$ transforms.
Additional software was used to create
meshes (gmsh~\cite{gmshCite}),
finite element data (Firedrake~\cite{Rathgeber2016}),
and renderings (ParaView~\cite{paraviewBook}). 
The Diderot code can be found in the supplementary materials.

%
%


\begin{figure}
\includegraphics[width=\columnwidth]{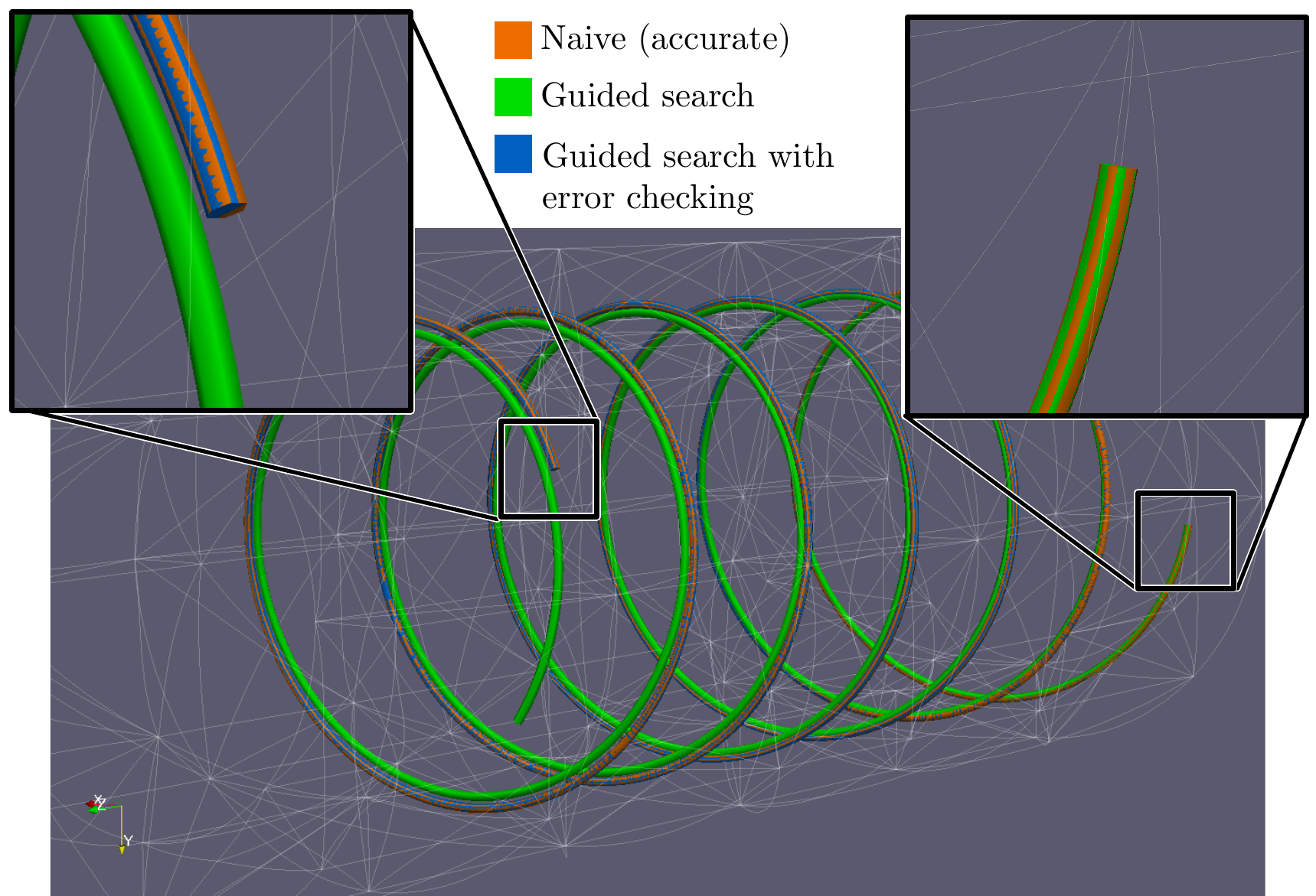}
\caption{\label{fig:stream} Streamlines in this synthetic vector
field in a curved mesh should be helices of constant radius. Three schemes
for updating position during integration are seeded at the
same location (right inset), but guided search (green) diverges
by the end (left inset). Guided search with error checking (blue)
very closely follows the accurate and more expensive naive method (orange).
}
\end{figure}

To test streamlines, we created a curved mesh
between two concentric cylinders.
We then interpolated $f(x,y,z) = (y, -x, 0.1)$ onto
a function space specified by the mesh and quadratic $P$.
An RK2 streamline program with
guided search produces the green path in Fig.~\ref{fig:stream}.
Also shown is an orange path produced
via the more expensive naive scheme, which shows the accuracy
of the blue path computed with the
error-checked guided search noted in \S\ref{sec:movement}.
This result is consistent with Coppola et al.'s accuracy analysis
of these schemes~\cite{coppola_nonlinear_2001}:
standard guided search errs in ways that
error-checked guided search avoids.

\begin{table}
  \caption{\label{fig:streamTable}
For computing streamlines with error-checked guided search (as in
Fig.~\ref{fig:stream}), over various step sizes (rows) and error
parameters (columns), the table gives run times in seconds and
speed-ups (in parentheses) of guided search relative to the naive
scheme.
Timing comparisons use equal numbers of steps within the mesh.
We note that the speedup results exhibit large variations between step sizes;
we hypothesize that the variations occur because decreasing step 
size can unpredictably both increase speedup via reducing error checking and 
decrease speedup via potentially increasing the number of points on a path
that are close to the initial guess of Newton's method, the elemental center, potentially improving the naive scheme's time~\cite{Blum1998}.
}
  \resizebox{\columnwidth}{!}{
\begin{tabular}{@{}r|llll@{}}
\toprule
& Error Parameter\hspace{-1.3em} & & & \\
Step Size & $10^{-4}$ & $10^{-5}$ & $10^{-6}$ &  \\ \midrule
0.2                       & 0.036s (2.524)            & 0.049s (2.220)            & 0.052s (2.056)            &  \\
0.02                      & 0.097s (9.183)            & 0.149s (6.735)            & 0.120s (7.846)            &  \\
0.002                     & 5.353s (2.600)            & 6.530s (2.369)            & 5.749s (2.423)            &  \\ \bottomrule
\end{tabular}}
\\

\end{table}

Coppola et al.\ also analyze the \emph{performance} of guided search.
Our preliminary results in Table~\ref{fig:streamTable} reproduce their
finding that error-checked guided search runs $2-10$ times faster
than the naive approach; this is notable considering that our code
is in a new high-level DSL.
We are also encouraged by this speed-up because it justifies the compiler and language
effort of \S\ref{sec:meshpos} and facilitates
future work on exploring new point movement techniques independently
of visualization methods.
\vspace{0.01cm} 
\begin{figure}
\includegraphics[width=\columnwidth]{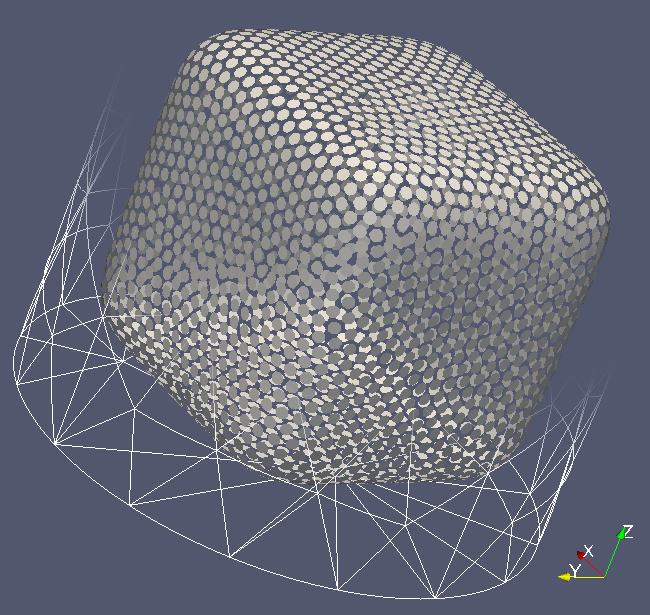}
\caption{\label{fig:iso} Particle based sampling of an isosurface
in the form of a rounded cube, contained with a cylindrical mesh,
the curved boundary of which is visible in the lower part.
}
\end{figure}

Meyer et al.\ pioneered isosurfacing via particle systems on curved geometries~\cite{meyer_particle_2007}.
It took us a few hours to adapt an $80$-line minimalist isosurface sampling
Diderot program~\cite{GLK:eurovis18} to produce Fig.~\ref{fig:iso},
showing a sampling of the isocontour $x^{6}+y^{6}+z^{6}=1$ in a cylindrical
mesh with hexic $P$.
Our result lacks curvature-adaptive sampling~\cite{meyer_particle_2007}, but it shows the
viability of our approach.

\begin{figure}
\includegraphics[width=\columnwidth]{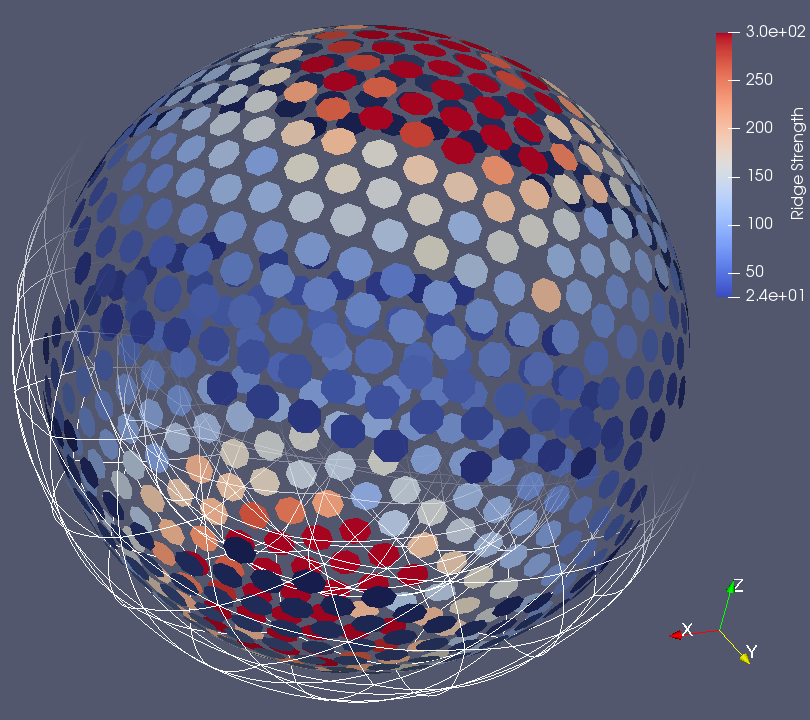}
\caption{\label{fig:ridg} Particle based sampling of a ridge surface on a curved geometry,
colormapped by feature strength.
The curved edges of boundary elements of the mesh are shown in lower part.
}
\end{figure}

We also sample a \emph{ridge surface} of the function
$f(x,y,z) = z^{2}\sin(x^{2}+y^{2}+z^{2})$, inspired by
Eberly's consideration of ridges in fluid flow
(c.f. Fig.~6.49 in~\cite{eberly_ridges_1996}). 
%
We created a curved mesh between two concentric spheres,
and approximated $f$ in a function space given by
the mesh and hexic $P$.
Since $f$ is non-polynomial, the approximation is at most
$C^{0}$ continuous across cell boundaries, which could
create discontinuities in the ridge surface itself.
A 300-line Diderot program for sampling general features with particle
systems was adapted as described in \S\ref{sec:meshpos},
and the results in Fig.~\ref{fig:ridg} were
found after experimenting with parameters
(feature strength threshold of 24, feature bias of 0.1).
We spent more time creating the example function and mesh
than we did writing and using the program.
%
Therefore, we feel that the most interesting aspect of this result
is in the DSL design.
The conceptual orthogonality of guided search and particle system
evolution was manifested as a clean separation in the code between
the implementation of those two methods, such that the guided
search code remained unchanged as the particle system
code was changed from isocontours to ridge surfaces.

\section{Discussion, Conclusions, and Future Work}
\label{sec:discuss}

Our methods and results demonstrate that point movement
and visualization algorithms can be orthogonal and composable.
We have shown how conceptually orthogonal algorithmic
components can be cleanly expressed as separate pieces of code
in a DSL. 
%
The ridge surface example additionally shows that this separation
extends in concept and in code to particle systems for a novel
visualization target for FEM data.
%
We hope these results convince readers of the potential of high-level
DSLs to create new visualization programs
by combining two separate specifications: one of the data
form (regular grid or FEM data), and one of the core visualization algorithm.
%

Directions of ongoing and future work are organized around
Diderot, guided search, and the expression of visualization
algorithms in general.
With respect to Diderot, we hope to add other data forms beyond regular grids
and finite element data such as Riemannian manifolds, where position
movement is given by solving a variational problem to find a geodesic.
With respect to guided search, we wish to augment it with information
from run-time or compile-time, and we wish to explore its application to
visualization methods that are less sensitive to errors in position
location (i.e., volume rendering or predictor-corrector schemes used in PVO line tracing~\cite{wkontopidis_predictor-corrector_1983,pagot_efficient_2011}).
%
%
Finally, we suspect that many existing implementations of visualization
algorithms currently specialized to a particular data form may contain ideas
that are as orthogonal and composable as guided search is for FEM data.
We hope to foster further research by uncovering those
ideas and exploring how they can be combined into new visualization algorithms,
implemented in idiomatic and re-usable code.
%



\acknowledgments{
This work was supported by NSF grant CCF-1564298.}

\bibliographystyle{abbrv}

\newpage
\section{
Supplementary Materials for Point Movement in a DSL for Higher-Order FEM Visualization
}

\section{Summary of Supplemental Materials}
\label{sec:sum}
In this document, we will include the commentated code that is used by
``Point Movement in a DSL for Higher-Order FEM Visualization.''
Each section will provide Diderot snippets that could be used to produce the figures in the paper.
These snippets are complete in the sense that they are either full programs in their own right
or could easily be combined with another snippet featured here to make a full program.
In addition, each section will provide some brief commentary on the code, highlighting both
the new ideas in the code not evident before and the work required to create these snippets
from older Diderot programs.
The snippets included below are:
\begin{enumerate}
\item A streamline program that uses guided search.
\item An overloaded function for the error checked guided search scheme.
\item An overloaded function for the naive scheme.
\item A small particle system used to create isosurfaces.
\item A larger particle system used to create ridge surfaces.
\end{enumerate}

Finally, we note that the figure captions provide information about parameters that we used
to create our figures if they are unstated in the paper.

\section{Streamlines}
\label{sec:stl}
\begin{figure*}[hbt]
\begin{minipage}{1.0\textwidth}
{\fontsize{6.2}{7.3}\selectfont
\begin{lstlisting}[xleftmargin=1.5em,multicols=2]
#version 3.0
//mesh specification:
type mesh mesh_t = file("evalProg.json");
//space (V) specification:
type functionSpace{mesh_t}[dim] fns_t = file("evalProg.json");
//u_{V} specification:
type femFunction{fns_t} func_t = file("evalProg.json"); 
const int dim = mesh_t.dim;        //dim of mesh

input mesh_t mesh; //input mesh
input fns_t space = fns_t(mesh);   //input space
input func_t data = func_t(space); //input u_{V}
refCell{mesh_t} K = mesh.refcell;  //get refCell

input int timeSteps=32;           //guided search step limit
input real timeEps = 0.0000001;   //1 - time max
//Overload + operator (see text for comments):
overload position{mesh_t} + (position{mesh_t} x,
                             tensor[dim] worldDelta) {
  if (!x.isValid){ return(x);}
  real time = 1; 
  position{mesh_t} cmp = x; 
  foreach (int i in 0..timeSteps ){
    tensor[dim, dim] iJac = inv($\nabla\otimes$(cmp.mc.transform)(cmp.refPos));
    tensor[dim] refDelta = iJac $\bullet$ worldDelta;
    tensor[dim] nPos = cmp.refPos + time * refDelta;
    if (K.isInside(nPos)) {
      return(cmp.mc.meshPos(nPos));
    } else { 
      tensor[dim] nRefDelta = normalize(refDelta);
      real eTime = K.exit(cmp, nRefDelta);
      time-=eTime / |refDelta|; 
      if(eTime == -1){
	return(mesh.findPos(x.worldPos() + worldDelta));
      } 
      position{mesh_t} nmp = K.exitPos(cmp, nRefDelta);
      if ( !nmp.isValid || time < timeEps){
	return(nmp);
      }
      cmp = nmp;
    }
  }
  return(mesh.findPos(x.worldPos() + worldDelta)); 
}
//evaluate u_{V} at a pos
function tensor[dim] nV(position{mesh_t} x){
  if(x.isValid()){                //border control
    cell{mesh_t} c = x.mc;        //get cell
    tensor[dim] ref = x.refPos(); //get refPos
    //f_{i}(y) where y = x.refPos()
    tensor[dim] val = data.funcCell(c).refField(ref);
    return(val/|val|);
  } else {return(zeros[dim]);}    //border control
}
input tensor[dim][] startPoints;//streamline start points
input real stepSize = 0.01;     //RK2 stepsize
input int stepMax = 32;         //RK2 steps max

strand line(tensor[dim] startPos){
  output tensor[dim][] stream = {};
  int step = 0;
  //find position corresponding to
  //global position of startPos
  position{mesh_t} cPos = mesh.findPos(startPos);
  update { //trace if position is valid
    if(!cPos.isValid()  || step == stepMax){stabilize;}
    stream = stream@{cPos.worldPos()};
    position{mesh_t} intermed = cPos + 0.5 * stepSize * nV(cPos);
    if(!intermed.isValid){stabilize;} //validity of substep
    cPos = cPos +  stepSize * nV(intermed); 
    step+=1;
  }
} create_collection {line(x) | x in startPoints};
\end{lstlisting}%
}%
\end{minipage}
\caption{\label{fig:stlgstex} A complete RK2 streamline program that uses guided search}
\end{figure*}

\begin{figure*}[hbt]
\begin{minipage}{1.0\textwidth}
{\fontsize{6.2}{7.3}\selectfont
\begin{lstlisting}[xleftmargin=1.5em,multicols=2]
input real errorMax = 1.0;
overload position{mesh_t} + (position{mesh_t} x,
                             tensor[dim] worldDelta) {
  if (!x.isValid){ return(x);}
  real time = 1;            
  position{mesh_t} cmp = x; 
  foreach (int i in 0..timeSteps ){
    tensor[dim, dim] iJac = inv($\nabla\otimes$(cmp.mc.transform)(cmp.refPos));
    tensor[dim] refDelta = iJac $\bullet$ worldDelta; 
    tensor[dim] nPos = cmp.refPos + time * refDelta;
    if (K.isInside(nPos)) { 
      position{mesh_t} nmp = cmp.mc.meshPos(newPos); 
      //check error:
      if(|nmp.worldPos() - (x.worldPos() + dPos) | > errorMax){
	vec3 guess = (cmp.worldPos() + time * dPos);
	if(cmp.mc.isInside(guess)){ //guess current cell
	  return(cmp.mc.meshPos(cmp.mc.inverseTransform(guess)));
	} else { //default to naive scheme
	  return(meshData.findPos(x.worldPos() + dPos));}
      }
      return(nmp);
    } else { 
      tensor[dim] nRefDelta = normalize(refDelta);
      real eTime = K.exit(cmp, nRefDelta);
      time-=eTime / |refDelta|; 
      if(eTime == -1){ 
	return(mesh.findPos(x.worldPos() + worldDelta));
      } 
      position{mesh_t} nmp = K.exitPos(cmp, nRefDelta);
      if ( !nmp.isValid ){return nmp;} //left the mesh
      vec3 truth =  (x.worldPos() + dPos * (1-time));
      real error = |nmp.worldPos() - truth|;
      if(error > errorMax){
	//if too much error, naive scheme:
	return(meshData.findPos( (x.worldPos() + dPos)));
      }
      if (time < timeEps){ //ran out of time.
	return(nmp); 
      }
      cmp = nmp;
    }
  }//spent too much time -  use naive scheme
  return(mesh.findPos(x.worldPos() + worldDelta)); 
}
\end{lstlisting}%
}%
\end{minipage}
\caption{\label{fig:stlgsectex} A guided search implementation that check errors. This code can be substituted into Figure~\ref{fig:stlgstex} in place of the addition overload to create a error checking guided search RK2 program.}
\end{figure*}

\begin{figure*}[hbt]
\begin{minipage}{1.0\textwidth}
  {
    \fontsize{6.2}{7.3}\selectfont
\begin{lstlisting}[xleftmargin=1.5em]
overload position{mesh_t} + (position{mesh_t} x,
			     tensor[dim] deltaPos)
{
  if (!x.isValid()){
    return(x);
  }
  return(meshData.findPos(x.worldPos() + dPos));
}

\end{lstlisting}%
}%
\end{minipage}
\caption{\label{fig:stls} An overloaded position operator that implements the naive scheme. This code can be substituted into Figure~\ref{fig:stlgstex} in place of the addition overload to create a naive scheme RK2 program.}
\end{figure*}

We provide line by line commentary on Figure~\ref{fig:stlgstex}, which features
a complete guided search streamline program.
Line 1 declares the version of the compiler that we use, which is 3.0.
Lines 2 through 7 declare the mesh, function space ($V$), and function types ($u_{V}$)
that the program uses.
The JSON files used in these declarations will
be documented further in future work, but we note that we automatically
generate them for Firedrake's meshes, spaces, and functions.

Lines 10 through 12 take inputs for the mesh, space, and function.
Line 13 provides the reference cell from the mesh.
Lines 18 through 44 are featured in the paper as the guided search algorithm.
Lines 46 through 54 supply an auxiliary function, nV, to unpack positions and evaluate
the field $f_{i}$ on the reference cell; this process consists in taking out a mesh cell, reference
position, and then using the information about $u_{V}$, the data, to get a field, which
is sampled at the reference position.
The auxiliary function checks the validity of the position and allows for some sort of
border control.
Lines 55 through 57 simply control the streamline algorithm, RK2.
Lines 58 though 73 are the streamline algorithm with some modifications to use positions:
First, line 64 takes the input vector to the strand, line, and converts it to a position.
Second, lines 66 and 69 provide border control by checking for the validity of the positions
(in older Diderot versions, this was done via an inside function.)
Finally, line 67 records the world position of a point on the streamline.
The core RK2 algorithm is legible in lines 68 and 70, which use the function
nV to sample a field at a position as in the older Diderot streamlines program.

We now roughly measure the changes required to turn a vanilla Diderot streamline
program to this program.
A standard Diderot streamline program would live in lines 46 through 73 after the
addition of image inputs and field declarations.
The changes in the function, nV, defined from 46 to 53, are changes in the field
evaluation and represent an additional 5 to 10 lines from the original program.
The strand definition is changed at lines 64 (type change and conversion), line
66 for checking the validity, line 67 for recording the world position, line 68
for a type change, and line 69 for another validity check.
Thus, the total changes to the streamline program amount to about 15 lines of code
besides the addition of the FEM inputs and guided search (line 1 through line 44)

In Figure~\ref{fig:stlgsectex}, we show the overloaded position operator with an
error checking functionality.
An error max parameter is required as a new input on line 1 and we now provide commentary
on its usage i.e the additions present in the error checked guided search.
The error parameter is used on lines 14 and 33 to check if the computed reference position
corresponds to a world space position that is close to the correct world space position
i.e the world space position of a naive position update.
If on either line 14 or 33, the condition is meet, the program continues as in the
previous figure, but otherwise, the new program tries to recover somehow.
On line 14, the code is finishing inside the reference cell, and, therefore, can
use the current cell to check if the correct world space position corresponds
to a reference position in the current cell.
If the current cell does contain the correct reference position, the correct
reference space position is computed via $T_{i}^{-1}$, but if the current cell
does contain the correct reference position, the naive scheme is used.
On line 33, the position is currently on a boundary so use of $T_{i}^{-1}$ is
inappropriate; thus, the naive scheme is used immediately on line 34.

Finally, in Figure~\ref{fig:stls}, we provide an overload for the naive scheme. This is provided for clarity and does not require comment.

\section{Particles}
\label{sec:part}

In this section, we will provide the complete programs used to make the particle system figures.
The position addition overload in these programs does not differ from Figure~\ref{fig:stlgstex}
nor do the creation of types or finite element inputs.
Ergo, in this section, we will focus on the changes to the core particle systems programs;
besides the addition of the overloaded functions and finite element inputs, how many lines
were changed in porting these programs?
How much did the main loop change?
We stress that one should examine the prior work on Diderot to understand
the full particle system programs and that here we mainly seek to point out
the limited extent to which FEM versions differ from the original programs.

\begin{figure*}[hbt]
\begin{minipage}{1.0\textwidth}
{\fontsize{6.2}{7.3}\selectfont
\begin{lstlisting}[xleftmargin=1.5em,multicols=2]
#version 3.0
//mesh specification:
type mesh mesh_t = file("evalProg.json");
//space (V) specification:
type functionSpace{mesh_t}[dim] fns_t = file("evalProg.json");
//u_{V} specification:
type femFunction{fns_t} func_t = file("evalProg.json"); 
const int dim = mesh_t.dim; //dim of mesh

input mesh_t mesh; //input mesh
input fns_t space = fns_t(mesh); //input space
input func_t data = func_t(space); //input u_{V}
refCell{mesh_t} K = mesh.refcell; //get refCell

input int timeSteps=32; //guided search step limit
input real timeEps = 0.0000001; //1 - time max
overload position{mesh_t} + (position{mesh_t} x,
                             tensor[dim] worldDelta) {
  if (!x.isValid){ return(x);}
  real time = 1; //time
  position{mesh_t} cmp = x; //current mesh pos
  foreach (int i in 0..timeSteps ){
    tensor[dim, dim] iJac = inv($\nabla\otimes$(cmp.mc.transform)(cmp.refPos));
    tensor[dim] refDelta = iJac $\bullet$ worldDelta; //reference velocity
    tensor[dim] nPos = cmp.refPos + time * refDelta;
    if (K.isInside(nPos)) {//inside K
      return(cmp.mc.meshPos(nPos)); //found x+v as (cmp.mc, nPos)
    } else { //we left the reference cell - compute when:
      tensor[dim] nRefDelta = normalize(refDelta);
      real eTime = K.exit(cmp, nRefDelta);
      time-=eTime / |refDelta|; //update time remaining.
      if(eTime == -1){//invalid direction, use naive scheme
	return(mesh.findPos(x.worldPos() + worldDelta));
      } //find the exit location in the next cell.
      position{mesh_t} nmp = K.exitPos(cmp, nRefDelta);
      if ( !nmp.isValid || time < timeEps){//left mesh or
	return(nmp); //ran out of time.
      }
      cmp = nmp;
    }
  }//spent too much time -  use naive scheme
  return(mesh.findPos(x.worldPos() + worldDelta)); 
}

//simple position subtraction operations
overload tensor[dim] - (position{mesh_t} x, position{mesh_t} y)
{
  if (x.isValid && y.isValid {
    return(x.worldPos() - y.worldPos());
  } else {return(zeros[dim]);}
}


//particle system controls
input real rad = 0.01;
input real eps = 0.01;
input real v0 = 0.0;
input tensor[dim][] ipos;


function tensor[dim] fStep(position{mesh_t} y){ 
  if(y.isValid){                              //border control
    tensor[dim]  x = y.refPos;                //get ref pos
    cell{mesh_t} c = y.mc;                    //get the cell
    cell{func_t} f = data.funcCell(c);        //get element info
    vec3 grad = $\nabla$(f.transformedRefField)(x);   //sample grad
    tensor[dim] ret = (v0 - (f.refField)(x)) * grad /(grad $\bullet$ grad);
    return ret;                               //return newton step
  } else {return [$\infty, \infty, \infty$];}                         //a big step
}

function tensor[dim, dim] fPerp(position{mesh_t} y){
  if (y.isValid){                              //border control
    tensor[dim] x = y.refPos;                  //get ref pos
    cell{mesh_t} c = y.mc;                     //get the cell
    cell{func_t} f = data.funcCell(c);         //get element info
    vec3 norm = normalize($\nabla$(f.transformedRefField)(x));
    return identity[dim] - norm $\otimes$ norm;         //return fPrep
   
  } else return(zeros[dim, dim]);
}
function real fStrength(position{mesh_t} y){
  if(y.isValid{                              //border control
      cell{mesh_t} c = y.mc;                 //get cell
      cell{func_t} f = data.funcCell(c);       //get element info
    return |$\nabla$(f.transformedRefField)((y.refPos))|;
  } else {return(0.0);}
}
//particle system core code:
function real phi(real r) = (1-r)^4;
function real phi'(real r) = -4 * (1-r)^3;
function real enr(tensor[dim] x) = phi(|x|/rad);
function tensor[dim] frc(tensor[dim] x) = phi'(|x|/rad) * (1/rad) * x/|x|;


strand particle(position{mesh_t}  pos0, real hh0){//changed the type
  output position{mesh_t} pos = pos0; //changed the type
  real hh = hh0;
  tensor[dim] step = zeros[dim];
  bool found = false;
  int nfs = 0;
  int test = 1;
  int testp = 0;
  update {
    if(!pos.isValid || fStrength(pos) == 0 || hh == 0){//add border control
      die;
    }
    if(!found) {
      step = fStep(pos);
      pos = pos + step;
      if(|step|/rad > eps){
	nfs += 1;
	if(nfs > 10) { die;}
      } else { found = true; testp=1;}
    }
    else {
	pos = pos + fStep(pos);
	step = zeros[dim];
	real oldE = 0;
	tensor[dim] force = zeros[dim];
	int nn = 0;

	foreach (particle P in sphere(rad)){
	  oldE += enr(P.pos - pos);
	  force += frc(P.pos - pos);
	  nn += 1;
	}
	if (0 == nn && pos.isValid()){ //added border control
	  new particle(pos + [0.5*rad, 0,0 ], hh);
	  continue;
	}
	force = fPerp(pos) $\bullet$ force;
	tensor[dim] es = hh*force;
	if(|es| > rad){
	  hh *= rad/|es|;
	  es *= rad/|es|;
	}
	position{mesh_t} samplePos = pos + es; //changed the type
	tensor[dim] fs = fStep(samplePos);
	if (|fs|/|es| > 0.5){
	  hh *= 0.5;
	  continue;
	}
        position{mesh_t} oldPos = pos;//changed the type
	pos += fs + es;
	real newE = sum {enr(pos - P.pos) | P in sphere(pos, rad)};
	if (newE - oldE > 0.5 * (pos - oldPos) $\bullet$ ( -force )) {
	  pos = oldPos;
	  hh *= 0.5;
	  continue;
	}
	  
	
	hh *= 1.1;
	step = fs + es;
	if (nn < 5){
	  new particle(pos + 0.5 * rad * normalize(es), hh);
	}
      }
    }
}
 update { //add an extra check (2 lines):
  bool allValid = all {P.pos.isValid | P in particle.all}; 
  bool allFound = all {P.found | P in particle.all};
  real maxStep = max {|P.step| | P in particle.all};
  if (allFound && allValid && maxStep/rad < eps) {stabilize;}
}
create_collection {particle(mesh.findPos(x), 1) | x in ipos};

\end{lstlisting}%
}%
\end{minipage}
\caption{\label{fig:iso} A complete though minimal particle system that uses guided search and is aimed at isosurfaces. We note that the figure in the paper used this program with eps=0.005, rad=0.5, and iso=0.0.}
\end{figure*}

We first consider the shorter particle systems program used to create the
isosurface in the paper, displayed in Figure~\ref{fig:iso}.
Lines 1 through 45 are basically identical to those for guided search in
Figure~\ref{fig:stlgstex}.
Line 46 through 52 implement position subtraction, which is a standard position
operation, via taking world space differences if both positions are valid,
and otherwise returning zero.
Lines 54 though 168 implement a small particle system for an isosurface.
Lines 54 through 58 are parameters to the system as in the original program.
Lines 61 through 89 implement feature strength, feature step,
and feature perpendicular functions that sample positions; these functions
are considered inputs to the particle system code.
These are different from the original versions of these functions,
but they are trivially the same only sampling through the reference
space via unpacking the position, accessing the current cells,
and acquiring the transformed reference fields described in the paper.
In terms of lines of code, each differs by 5 to 10 lines from the original,
leading to an additional roughly 10 to 20 lines, but, we note these functions
live outside the core particle system code.
The core of the particle systems program lies in lines 89 through
168.
We observe that this is basically identical to the original program except
for 7 lines (lines 96,97,105, 128, 144, 163, 166).
Each line has a comment explaining the change relative to the old program,
but we note that 4 lines differ by type, two lines use the validity method
of a position to do border control, and only lines 163 and 166 implements new functionality.
In short, for a core loop of 72 lines, only 2 to 4 were added or changed non-trivially.
In short, the core logic of the small particle system program is basically unchanged
in the conversion to use guided search modulo the addition of the overloads on
positions and the specification of finite element data.
Examining our analysis, we see that besides the FEM inputs and position overloads, the
program features at most 30 lines of additions or changes.

\begin{figure*}[b]
{\fontsize{6.2}{7.3}\selectfont
\begin{lstlisting}[xleftmargin=1.5em,multicols=2]
#version 3.0
//mesh specification:
type mesh mesh_t = file("evalProg.json");
//space (V) specification:
type functionSpace{mesh_t}[dim] fns_t = file("evalProg.json");
//u_{V} specification:
type femFunction{fns_t} func_t = file("evalProg.json"); 
const int dim = mesh_t.dim;        //dim of mesh

input mesh_t mesh; //input mesh
input fns_t space = fns_t(mesh);   //input space
input func_t data = func_t(space); //input u_{V}
refCell{mesh_t} K = mesh.refcell;  //get refCell

input int timeSteps=32;           //guided search step limit
input real timeEps = 0.0000001;   //1 - time max
//Overload + operator (see text for comments):
overload position{mesh_t} + (position{mesh_t} x,
                             tensor[dim] worldDelta) {
  if (!x.isValid){ return(x);}
  real time = 1; 
  position{mesh_t} cmp = x; 
  foreach (int i in 0..timeSteps ){
    tensor[dim, dim] iJac = inv($\nabla\otimes$(cmp.mc.transform)(cmp.refPos));
    tensor[dim] refDelta = iJac $\bullet$ worldDelta;
    tensor[dim] nPos = cmp.refPos + time * refDelta;
    if (K.isInside(nPos)) {
      return(cmp.mc.meshPos(nPos));
    } else { 
      tensor[dim] nRefDelta = normalize(refDelta);
      real eTime = K.exit(cmp, nRefDelta);
      time-=eTime / |refDelta|; 
      if(eTime == -1){
	return(mesh.findPos(x.worldPos() + worldDelta));
      } 
      position{mesh_t} nmp = K.exitPos(cmp, nRefDelta);
      if ( !nmp.isValid || time < timeEps){
	return(nmp);
      }
      cmp = nmp;
    }
  }
  return(mesh.findPos(x.worldPos() + worldDelta)); 
}

//Overload - operator
overload tensor[dim] - (position{mesh_t} x, position{mesh_t} y)
{
  if (x.isValid && y.isValid {
    return(x.worldPos() - y.worldPos());
  } else {return(zeros[dim]);}
}
//particle system controls
input real fStrTh ("Feature strength threshold");
input real fMaskTh ("feature mask threshold") = 0;
input real fBias ("Bias in feature strength computing") = 0.0;
input real tipd ("Target inter-particle distance") = 1.0;/*@\label{line:pbfs-tipd}@*/
input real mabd ("Min allowed birth distance (> 0.7351)") = 0.75;/*@\label{line:pbfs-mabd}@*/
input real travMax ("Max allowed travel to or on feature") = 10;
input int nfsMax ("Max allowed # feature steps ") = 20;
input real gdeTest ("Scaling in sufficient decrease test") = 0.5;
input real gdeBack ("How to scale stepsize for backtrack") = 0.5;
input real gdeOppor ("Opportunistic stepsize increase") = 1.2;
input real fsEps ("Conv. thresh. on feature step size");
input real geoEps ("Conv. thresh. on system geometry") = 0.1;
input real mvmtEps ("Conv. thresh. on point movement") = 0.01;
input real rpcEps ("Conv. thresh. on recent pop. changes") = 0.01;
input real pcmvEps ("Motion limit before PC") = 0.3;
input real isoval ("Which isosurface to sample") = 0;
input int verb ("Verbosity level") = 0;
input real sfs ("Scaling (<=1 for stability) on fStep") = 0.5;
input real hist ("How history matters for convergence") = 0.5;
input int pcp ("periodicity of population control (PC)") = 5;
input vec3[] ipos ("Initial point positions");
input int fDim = 2;

//Freature Functions:
function vec3 fStep(position{mesh_t} pos) {
  if(pos.isValid){
    vec3 x = pos.refPos;
    cell{mesh_t} c = pos.mc;
    cell{func_t} f = data.funcCell(c);
    vec3 g = $\nabla$(f.transformedRefField)(x);
    tensor[dim, dim] H = $\nabla\otimes\nabla$(f.transformedRefField)(x);
    vec3[3] E = evecs(H);
    real[3] L = evals(H);
    vec3 up =  -(1/L[2])*E[2]$\otimes$E[2]•g;
    return up;
  } else {return([$\infty, \infty, \infty$]);}
}


function tensor[3,3] fPerp(position{mesh_t} pos) {
  if(pos.isValid){
    vec3 x = pos.refPos;
    cell{mesh_t} c = pos.mc;
    cell{func_t} f = data.funcCell(c);
    tensor[dim, dim] H = $\nabla\otimes\nabla$(f.transformedRefField)(x);
    vec3 E2 = evecs(H)[2];
    mat3 m =  identity[3] - E2$\otimes$E2;
    return m;
  } else {return(zeros[dim, dim]);}
}

function real fStrength(position{mesh_t} pos) {
  if(pos.isValid){
    vec3 x = pos.refPos;
    cell{mesh_t} c = pos.mc;
    cell{func_t} f = data.funcCell(c);
    vec3 g = $\nabla$(f.transformedRefField)(x);
    tensor[dim, dim] H = $\nabla\otimes\nabla$(f.transformedRefField)(x);
    real str = -evals(H)[2]/(fBias + |g|);
    return str;
  } else {return(0.0);}
}

function real fMask(position{mesh_t} x) = 0.0;
function bool fTest(position{mesh_t} x) = true;

function bool posTest(position{mesh_t} x) =
     (x.isValid               // Valid test replaces border control
     && fStrength(x) > fStrTh // possibly near feature
     && fMask(x) >= fMaskTh   // meets feature mask
     && fTest(x));            // passes addtl feature criterion
//End feature functions
//Auxiliary Particle system stuff:
int nnmin = 6 if (2==fDim) else 2 if (1==fDim) else 0;
int nnmax = 8 if (2==fDim) else 3 if (1==fDim) else 0;

function real phi(real r) {/*@\label{line:pbfs-phi}@*/
  real s=r-2.0/3;
  return
    1 + r*(-5.646 + r*(11.9835 + r*(-11.3535 + 4.0550625*r)))
  if r < 2.0/3 else
    -0.001 + ((0.09 + (-0.54 + (1.215 - 0.972*s)*s)*s)*s)*s
  if r < 1 else 0;
}
function real phi'(real r) {
  real t=3*r-2;
  return
    -5.646 + r*(23.967 + r*(-34.0605 + 16.22025*r))
  if r < 2.0/3 else
    0.01234567901*t*(4.86 + t*(-14.58 + t*(14.58 - 4.86*t)))
  if r < 1 else 0;
}
real phiWellRad = 2/3.0;    
real rad = tipd/phiWellRad; 
function real enr(vec3 x) = phi(|x|/rad);/*@\label{line:pbfs-enr}@*/
function vec3 frc(vec3 x) = phi'(|x|/rad) * (1/rad) * x/|x|;/*@\label{line:pbfs-frc}@*/


real pchist = hist^(1.0/(2*pcp));

int iter = 0;        
real rpc = 1;        
int popLast = -1;    

function real urnd(real x) {/*@\label{line:pbfs-urnd}@*/
  if (x==0) return 0;
  real l2 = log2(|x|);
  real frxp = 2^(l2-floor(l2)-1);
  return fmod((2^20 + 2*iter)*frxp, 1);
}


function real v3rnd(position{mesh_t} p){ //type change
  if(!p.isValid){                        //border control
    return 0;
  }
  vec3 v = p.refPos;                     //get ref pos
  return(fmod(urnd(v[0]) + urnd(v[1]) + urnd(v[2]), 1));
}

//Another type change:
function real genID(position{mesh_t} v) = floor(1000000*v3rnd(v));/*@\label{line:pbfs-genid}@*/

function int pcIter(){
  if (pcp>0 && iter>0 && 0 == iter % pcp) {return (((iter/pcp)%2)*2 - 1);}
  else {return 0;}
}
\end{lstlisting}%
}%
\caption{\label{fig:ridg1} Part 1 of a particle system that uses guided search and is aimed at ridge surfaces. This section contains the search and the particle system parameters. The parameters used to produce the figure in the paper are fStrTh=24, fBias=0.1, tipd==0.1, fsEps=geoEps=mvmtEps=0.1, rpcEps=0.01, pcmvEps=0.3, sfs=hist=0.5, pcp=5.}
\end{figure*}

\begin{figure*}[b]
{\fontsize{6.2}{7.3}\selectfont
\begin{lstlisting}[xleftmargin=1.5em,multicols=2]
strand point (position{mesh_t} p0, real hh0) { //changed the type
  output position{mesh_t} pos = p0; // changed the type
  real ID = genID(p0);  
  real hh = hh0;        
  vec3 step = [0,0,0];  
  bool found = false;   
  int nfs = 0;          
  real trav = 0;        
  real mvmt = 1;        
  real closest = rad;   
  int born = 0;         
  bool first = true;    
  update {
    if (!posTest(pos)) {
      die;
    }
    if (travMax > 0 && trav > travMax) {
      die;
    }
    if (!found) {
      if (nfsMax > 0 && nfs > nfsMax) {
        die;
      }
      step = sfs*fStep(pos);
      pos = pos +  step;
      mvmt = lerp(|step|/tipd, mvmt, hist);
      if (mvmt > fsEps) {
        trav += |step|/tipd;
        nfs += 1;
      } else {
        found = true;
        mvmt = 1;
        trav = 0;
      }
    } else { 
      
      if (0 == fDim) { stabilize; }
      step = sfs*fStep(pos); pos = pos + step; trav += |step|/tipd;
      real oldE = 0;             
      vec3 force = [0,0,0];      
      int nn = 0;                
      foreach (point P in sphere(rad)) {
        vec3 off = P.pos - pos;
        if (|off|/tipd < fsEps && ID <= P.ID) {
          die;
        }
        oldE += enr(off);
        force += frc(off);
        nn += 1;
      }
      if (0 == nn) { 
        if (!( pcIter() > 0 && born < nnmax )) { continue; }
        vec3 noff0 = fPerp(pos)$\bullet$[tipd,0,0];
        vec3 noff1 = fPerp(pos)$\bullet$[0,tipd,0];
        vec3 noff2 = fPerp(pos)$\bullet$[0,0,tipd];
        vec3 noff = noff0;
        noff = noff if |noff| > |noff1| else noff1;
        noff = noff if |noff| > |noff2| else noff2;
	//changed the type:
        position{mesh_t} npos = pos + tipd*normalize(noff); 
        npos = npos +  sfs*fStep(npos);
        if (posTest(pos)) {
          new point(npos, hh); born += 1;
        }
        continue;
      }
      vec3 es = hh*fPerp(pos)•force; 
      if (|es| > tipd) {             
        hh *= tipd/|es|;             
        es *= tipd/|es|;
      }                              
      vec3 fs = sfs*fStep(pos+es);   
      if (|fs|/(fsEps*tipd + |es|) > 0.5) {
        hh *= 0.5; 
        continue;
      }
      //changed the type:
      position{mesh_t} oldpos = pos;
      vec3 up = fs + es;
      pos = pos + up;
      real newE = 0;
      closest = rad;

      vec3 mno = [0,0,0];
      nn = 0;
      foreach (point P in sphere(rad)) {
        vec3 off = P.pos - pos;
        newE += enr(off);
        closest = min(closest, |off|);
        mno += off;
        nn += 1;
      }
      mno /= nn;

      if (newE - oldE > gdeTest*(pos - oldpos)$\bullet$(-force)) {
        hh *= gdeBack;
        if (0 == hh) {
          die;
        }
        pos = oldpos;
        continue;
      }
      hh *= gdeOppor;
      step += fs + es;
      trav += |step|/tipd;
      mvmt = lerp(|step|/tipd, mvmt, hist);
      if (|step|/tipd < pcmvEps && pcIter() != 0) {
        if (pcIter()>0      
            && newE<0       
            && nn<nnmin     
            && born<nnmax) {//changed the type:
          position{mesh_t} npos = pos + (-tipd*normalize(mno));
          npos = npos + sfs*fStep(npos); npos = npos + sfs*fStep(npos);
          bool birth = true;
          if (fDim == 2 && nn >= 4) {
            foreach (point P in sphere(npos, tipd*mabd)) {
              birth = false;
            }
            if (birth) {
              birth = v3rnd(pos) < (nnmin - nn)/real(nnmin);
            }
          }
          if (birth && posTest(npos)) {
            new point(npos, hh); born += 1;
          }
        } else if (pcIter() < 0 && newE > 0 && nn > nnmax) {
          if (v3rnd(pos) < (nn - nnmax)/real(nn)) {
            die;
          }
        }
      }
    } 
    first  = false;
  }
}
update {/*@\label{line:pbfs-global}@*/
  int pop = numActive();
  int pc = 1 if pop != popLast else 0;
  rpc = lerp(pc, rpc, pchist);
  bool allfound = all { P.found | P in point.all};
  real percfound =
    100* mean { 1.0 if P.found else 0.0 | P in point.all};
  real meancl = mean { P.closest | P in point.all };
  real varicl = mean { (P.closest - meancl)^2 | P in point.all };
  real covcl = sqrt(varicl) / meancl;
  real maxmvmt = max { P.mvmt | P in point.all };
  //added new convergence test
  //to find avoid saving invalid positions:
  bool allValid = all {P.pos.isValid | P in point.all};
  if (allfound            
      && covcl < geoEps   
      && maxmvmt < mvmtEps
      && rpc < rpcEps && allValid) { // use this new test.
    stabilize;
  }

  iter += 1;
  popLast = pop;
}
create_collection { point(mesh.findPos(p), 1) | p in ipos};
  
\end{lstlisting}%
}%
\caption{\label{fig:ridg2} Part 2 of a particle system that uses guided search and is aimed at ridge surfaces. This section contains the core of a Diderot program: the strand definition.}
\end{figure*}

This result is repeated with the larger particle system program, which is featured
for ridge surfaces in Figure~\ref{fig:ridg1} and Figure~\ref{fig:ridg2}.
Combining these figures yields the full program.
In the first part of the code, Figure~\ref{fig:ridg1}, we find the particle
system controls, guided search controls, and particle system auxiliary functions.
Lines 1 through 53 provide FEM inputs and guided search, as in the previous examples.
Lines 54 though 75 provide the particle system controls.
Lines 78 through 126 provide the feature step, perpendicular, strength, mask, and test functions
that these particle systems take as inputs.
As before, these are the same as their vanilla versions, but they take positions,
unpack them, and sample from the reference cell.
We don't provide additional commentary on these functions as the conceptual extent to which
they differ from the original Diderot programs is the same as in the previous particle
system.
The step, perpendicular, and strength functions require a few additional lines (at most 5 each)
where as the mask, test, and posTest functions change by 1 line, 1 line, and 2 lines
respectively.
Thus lines 78 through 126 feature at most 20 additions or changes to the code.
Lines 127 through 180 are almost identical to the original code, but the v3rand
and genID functions (lines 175 and 166) need to use positions instead of vectors,
creating another 4 changes.
Moving on to the second part of the large particle system, Figure~\ref{fig:ridg2},
we find the core particle system code.
In Figure~\ref{fig:ridg2}, all comments have been removed except those that indicate
that a change has been made from the original program.
We find that there are no more than 7 lines of changes and that
only the last two lines, which check that all positions are valid before allowing
convergence, are substantial changes.
Combining this with our analysis of the first part and, as before, discounting
the added overloads and FEM inputs, the total changes between this program and
its regular grid version amount to fewer than 35 lines of code.

\end{document}